\documentclass[twocolumn,longbibliography,aps,prx,floatfix,
preprintnumbers]{revtex4-2}

\usepackage[normalem]{ulem}
\usepackage{graphicx}
\usepackage{bm}
\usepackage{color}
\usepackage{xcolor}
\usepackage{amsmath}
\usepackage{amssymb}
\usepackage[T1]{fontenc}
\usepackage{lipsum}
\usepackage{gensymb}

\usepackage{multirow}

\usepackage{scrextend}
\usepackage[urlcolor=blue,colorlinks=true,citecolor=blue,linkcolor=blue,pdfstartview={FitH},bookmarks=false]{hyperref}

\begin{document}

\title{Magnetic and electrical transport properties of the single-crystalline half-Heusler antiferromagnet DyNiSb}
    \author{Abhinav Agarwal}
    \author{Prabuddha Kant Mishra}
    \author{Orest Pavlosiuk}
    \author{Maciej J. Winiarski}
    \author{Piotr Wiśniewski} 
    \author{Dariusz Kaczorowski} 
    \email{d.kaczorowski@intibs.pl}

    \affiliation{Institute of Low Temperature and Structure Research, Polish Academy of Sciences, Okólna 2, 50-422 Wrocław, Poland}

\begin{abstract}
High-quality single crystals of the half-Heusler compound DyNiSb were investigated for their low-temperature thermodynamic and magnetotransport properties. Magnetic susceptibility, heat capacity and electrical resistivity measurements revealed two distinct magnetic phase transitions at $T_\mathrm{{N1}}$ = 7.3~K and $T_\mathrm{{N2}}$ = 3.4~K, contrasting with previous reports on polycrystalline samples, which identified only a single transition near $T_\mathrm{{N2}}$. Moreover, the studied samples were found to exhibit metallic-like conductivity, at odds with a semiconducting behavior reported for the polycrystals. Magnetoresistance measurements performed in both transverse and longitudinal configurations revealed in small magnetic fields a weak antilocalization effect that diminishes with increasing temperature, giving way to a positive, monotonic magnetoresistance at high temperatures. Angular-dependent resistivity studies showed a crossover from four-fold to two-fold symmetry with increasing magnetic field strength, suggesting a field-induced reconstruction of the Fermi surface. Our findings highlight a complex magnetic and electronic transport behavior in DyNiSb, highly sensitive to structural disorder and easily tunable by external magnetic field.
\end{abstract}

\maketitle

\section{Introduction}
Half-Heusler (HH) compounds have attracted significant attention from the materials science community due to their tunable physical properties, including high thermoelectric efficiency \cite{Mastronardi1999, Bos2014, Fu2015, Ciesielski2020, Ciesielski} and non-trivial topological characteristics \cite{Chadov2010, Liu2011, Nowak2015, Guo2018}. These compounds typically crystallize with a noncentrosymmetric face-centered cubic lattice of the MgAgAs-type (space group $F\bar{4}3m$). They bear a general chemical formula XYZ, where X and Y are usually transition metals (TM) or rare-earth (RE) elements, and Z is a main-group element from the $4p$ or $5p$ block. 

HH compounds are well known for intrinsic structural imperfections \cite{Xie2014, Harmening2009, Ciesielskid}. Subtle variations in atomic site occupancy or antisite disorder can lead to significant changes in their electronic structures, with behavior ranging from semiconducting to semimetallic or metallic \cite{Larson2000}. Consequently, previous studies on polycrystalline samples have often reported inconsistent electrical transport properties. For instance, early experimental work on the compound DyNiSb showed semiconducting-like resistivity in the entire temperature range examined \cite{skolozdra1997}, while the subsequent investigations reported a non-monotonic temperature variation of the resistivity with a broad maximum near 150 K \cite{karla1998, Pierre2000}. In contrast, recent studies found metallic-type properties on similar RE-based HH antimonides \cite{Ciesielski2020}. These discrepancies can be attributed to a narrow-gap semiconducting nature of the HH compounds \cite{Winiarski2019}, where structural disorder of various kind may introduce a finite density of states at the Fermi level \cite{Pavlosiuk2021}, giving rise to a wide range of electrical transport behavior. 

Several members of the RENiSb family were extensively studied in polycrystalline form and considered promising candidates for thermoelectric applications \cite{Ciesielski, Ciesielski2020}. However, to date, no comprehensive investigations have been conducted on single crystals of the RENiSb compounds, leaving a gap in the understanding of the intrinsic behavior in these materials. Here we report the results of our research on the thermodynamic and magnetotransport properties of single-crystalline DyNiSb, examined in wide ranges of temperature and magnetic field. Magnetic measurements were performed in an external magnetic field applied along the [001] direction in the crystallographic unit cell of DyNiSb. The magnetoresistance was studied in both transverse and longitudinal configurations of electric current and magnetic field, which was directed at various angles relative to the [001] direction. To investigate the impact of various types of disorder on the electronic band structure, the density functional theory (DFT) calculations were performed.

\section{Experimental Details}
Single crystals of DyNiSb were synthesized using a two-step process. Initially, a polycrystalline ingot was prepared by arc melting the high-purity elemental constituents Dy, Ni, and Sb under a Ti-gettered argon atmosphere. To compensate for possible losses during the melting process, we added a slight (5\%) excess of Sb. To ensure homogeneity, the ingot was flipped over and remelted several times.

The obtained polycrystalline sample was then ground into fine powder using an agate mortar. This powder was mixed with Bi flux in a molar ratio of 1:30. The charge was placed in an alumina crucible and sealed inside an evacuated quartz ampule. The tube was heated to 1050$^{\circ}$C over 20 hours and held at this temperature for 24 hours to ensure complete homogenization. Following this, the temperature was slowly reduced to 650$^{\circ}$C at a controlled rate of 2$^{\circ}$C per hour. Subsequently, the ampule was quickly removed from the hot furnace and centrifuged to separate crystals from the flux.
 
Chemical composition of the obtained crystals was verified using energy-dispersive X-ray spectroscopy (EDS) performed using a FEI scanning electron microscope equipped with a Genesis XM4 EDS probe. The EDS analysis yielded the stoichiometry Dy:Ni:Sb=35:33:32 that is fairly close to the ideal equiatomic composition.

\begin{figure}
\includegraphics[width= \linewidth]{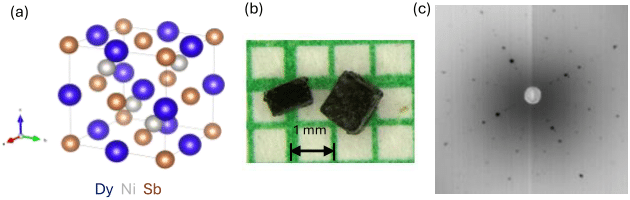}
 \caption{(a) Crystallographic unit cell of DyNiSb. (b) Photograph of as-grown single crystals of DyNiSb. (c) Laue diffraction pattern of DyNiSb single crystal, collected for the (001) plane.}
\label{crystal}
\end{figure}

The DyNiSb compound crystallizes with the MgAgAs-type unit cell (see Fig.~\ref{crystal}(a)), where the constituent atoms occupy the following Wyckoff positions: Dy at 4$a$ (0, 0, 0), Sb at 4$b$ (0.5, 0.5, 0.5), and Ni at 4$c$ (0.25, 0.25, 0.25) \cite{Ciesielski2020}. The single crystals grown in this work were of millimeter-size and had well-defined facets and shiny surfaces (see Fig.~\ref{crystal}(b)). Their crystalline quality and their crystallographic orientation were checked by backscattering Laue X-ray diffraction technique implemented in a Proto Manufacturing Laue-COS system, as exemplified in Fig.~\ref{crystal}(c). 

Magnetic measurements were carried out in the temperature interval 2-300\,K in magnetic fields up to 7\,T using a Quantum Design MPMS-XL superconducting quantum interference device (SQUID) magnetometer. The heat capacity was measured from 2 to 300\,K employing a relaxation technique implemented in a Quantum Design PPMS-9 platform. In all the thermodynamic studies, the external magnetic field was applied along the [001] direction. Magnetotransport measurements were performed over the temperature interval 2-300 K in magnetic fields up to 7 T using a standard four-probe AC technique and the same PPMS equipment. Electrical contacts were prepared on the (001) plane using silver wires and silver paint. The measurements were done in magnetic fields directed at various angles with respect to the electric current, which was applied along [010] crystallographic direction.

The density functional theory (DFT) calculations were done with the VASP package \cite{VASP1,VASP2,VASP3}. The generalized gradient parameterization (GGA \cite{GGA}) of the exchange-correlation functional was selected. The 600 eV plane-wave cut-off was used. The Dy $4f$ states were treated as core states. The calculations of defect formation energies were carried out in the scalar relativistic approach with 96-atomic supercells, whereas the fully relativistic band structures were obtained for conventional unit cells of the HH phase. Full geometry optimization was performed for all systems. The 6$\times$6$\times$6 and 12$\times$12$\times$12 {\bf k}-point meshes were employed for supercells and unit cells, respectively. The point defect formation energies at 0\,K were calculated \cite{defect_theory} according to the formulas:
\begin{equation}
E(V_X) = E_{defect} - E_{ideal} + \mu_X ,
\end{equation}
\begin{equation}
E(I_X) = E_{defect} - E_{ideal} - \mu_X ,
\end{equation}
\begin{equation}
E(X_Y) = E_{defect} - E_{ideal} + \mu_Y - \mu_X ,
\end{equation}

\noindent where $V_X$, $I_X$, and $X_Y$ are vacancy, interstitial, and antisite (X ion occupying Y ion position) defects, respectively. $E_{defect}$ and $E_{ideal}$ denote the total energies of particular structures with and without defects, respectively. Chemical potentials $\mu$ are related to the total energies per ion in elemental solids. A similar methodology was recently applied for other HH materials \cite{hH_defect}.

\section{RESULTS AND DISCUSSION}

\subsection{Magnetic properties and heat capacity}
\begin{figure}
\includegraphics[width= \linewidth]{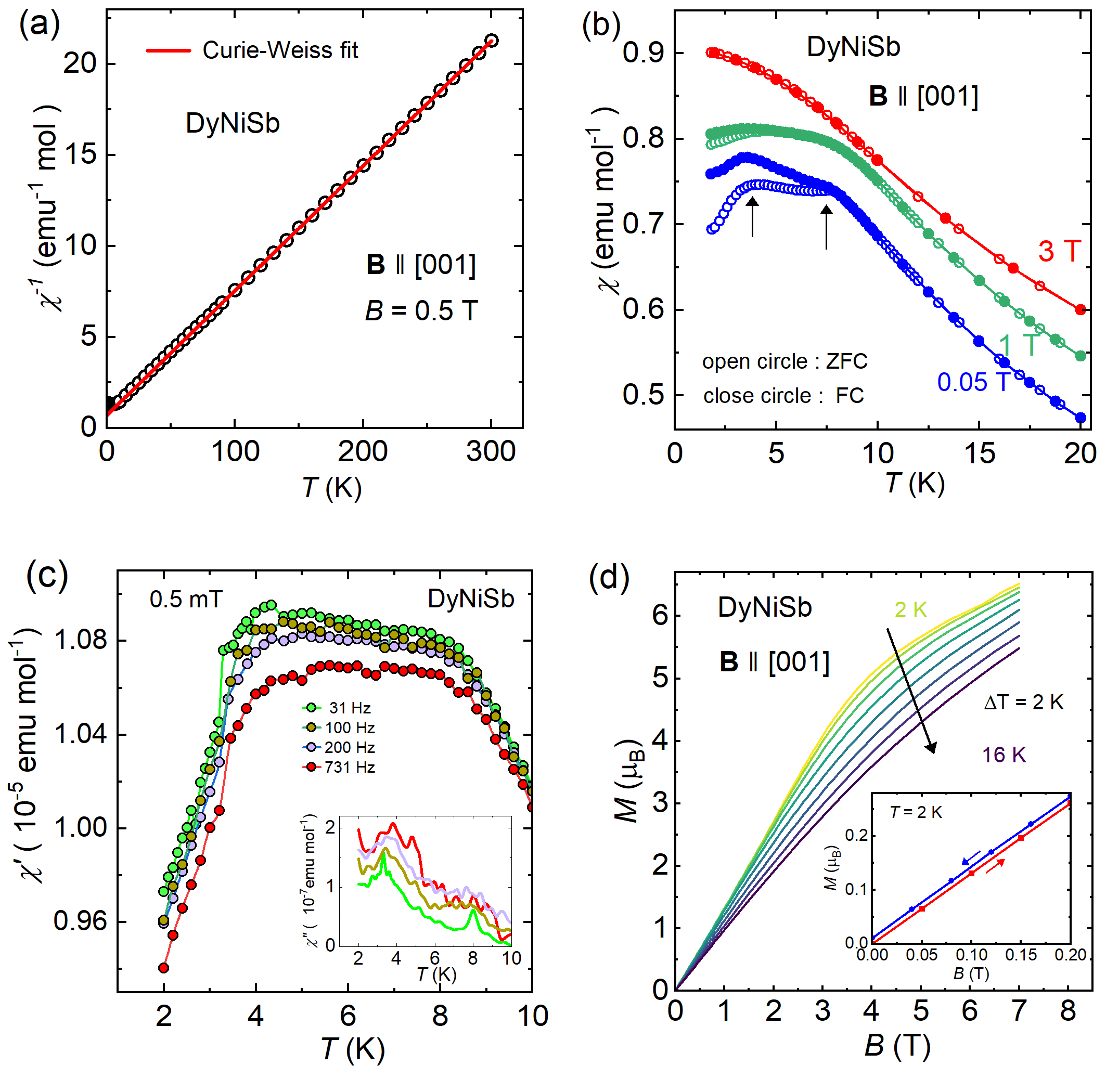}
\caption{(a) Temperature dependence of the reciprocal molar magnetic susceptibility of single-crystalline DyNiSb measured in a magnetic field of 0.5 T applied along the [001] direction. The solid red line represents the Curie-Weiss fit described in the text. (b) Low-temperature variations of the magnetic susceptibility of DyNiSb measured along the [001] direction in different magnetic fields upon cooling the sample in zero (ZFC) and applied (FC) field. Black arrows mark the transition temperatures. For a sake of clarity, the data obtained in $B=$ 1 T and 3 T were shifted upwards by offset values of 0.07 and 0.14 emu\,mol$^{-1}$Oe$^{-1}$, respectively. (c) Temperature dependence of the real part of the dynamical susceptibility measured in zero steady magnetic field in an ac field of 0.5~mT alternating with different frequencies. The inset presents the the temperature variation of the imaginary part of the ac magnetic susceptibility taken in the same conditions. (d) Magnetization isotherms measured for single-crystalline DyNiSb at various temperatures in magnetic fields applied along the [001] axis. The inset shows the data taken at $T=$ 2~K with increasing and decreasing the magnetic field strength.}
\label{mag}
\end{figure}

Fig.~\ref{mag}(a) presents the temperature variation of the inverse magnetic susceptibility, $\chi^{-1}(T)$, of DyNiSb, measured in an external magnetic field $\textbf{B}$ of 0.5~T applied along the crystallographic [001] direction. As can be inferred from this figure, the experimental data can be well described by the Curie-Weiss law (the least-squares fit was made above 100~K to avoid possible crystalline electric field effect). The so-obtained effective magnetic moment $\mu_\mathrm{eff}$ = 10.4 $\mu_\mathrm{B}$ is close to the theoretically predicted $\mu_\mathrm{eff,th}$ = 10.65 $\mu_\mathrm{B}$ of a free Dy$^{3+}$ ion. In turn, the paramagnetic Curie temperature $\theta_\mathrm{p}$ equals $-10.2$~K; its negative sign suggests that magnetic exchange interactions between the Dy$^{3+}$ spins are predominantly antiferromagnetic (AFM) in character. It should be noted that the obtained $\mu_\mathrm {eff}$ and $\theta_\mathrm{p}$ parameters have values similar to those reported for polycrystalline samples of DyNiSb \cite{Karla}. 

In order to check for long-range magnetic ordering, the temperature dependencies of the DC magnetic susceptibility of DyNiSb single crystals were measured at low temperatures in different applied magnetic fields ($\textbf{B} \parallel [001]$) using zero-field-cooling (ZFC) and field-cooling (FC) protocols. As displayed in Fig. \ref{mag}(b), in a field of 0.05~T, two clear anomalies occur in $\chi(T)$, which both bear an AFM character. The transition temperatures are equal to $T_\mathrm{N1}$ = 7.3~K and $T_\mathrm{N2}$ = 3.4~K. This result differs from that obtained in the measurements of polycrystalline DyNiSb, which indicated a single AFM transition at 3.5~K \cite{Karla, synoradzki2018}. Remarkably, the ZFC and FC curves bifurcate below $T_\mathrm{N1}$, suggesting some contribution of weak ferromagnetic component. However, as the applied field increases, the bifurcation quickly diminishes. At the same time, both transitions smear out, $T_\mathrm{N1}$ systematically decreases, in a manner characteristic of antiferromagnets, while $T_\mathrm{N2}$ remains nearly unaffected at least up to 1~T. In $B$ = 3~T, both transitions are suppressed, and the ZFC and FC curves fully overlap. 

In order to further examine the low-temperature magnetic behavior in DyNiSb, ac magnetic susceptibility measurements were performed. As can be inferred from Fig.~\ref{mag}(c), the real part of the ac susceptibility, $\chi^{\prime}(T)$ clearly reveals two AFM transitions at  $T_\mathrm{N1}$ and $T_\mathrm{N2}$ and shows no significant frequency dependence. In turn, the imaginary component $\chi^{{\prime}{\prime}}(T)$ shows a small upturn below $T_\mathrm{N1}$, in line with the proposed scenario of the presence of weak ferromagnetic component in the overall AFM structure of the compound. Verification of the actual magnetic structures of DyNiSb below $T_\mathrm{N1}$ and $T_\mathrm{N2}$ requires further studies, preferably by means of neutron diffraction.

Fig.~\ref{mag}(d) displays the field variations of the magnetization, $M(B)$, measured at low temperatures with a temperature interval $\Delta T$ = 2~K in magnetic fields $\textbf{B}\parallel$ [001]. The isotherms taken at the lowest temperatures show faint inflections near 3~T, which can be attributed to a metamagnetic transition, as expected for AFM materials. In stronger fields, they bend towards saturation, however fully polarized state is not achieved up to the strongest field applied. In $B =$ 7~T, the magnetization measured at $T =$ 2~K attains a value of about 6.5~$\mu_\mathrm{B}$/f.u., which is significantly smaller than $M_\mathrm{s,th}$ = 10.0 $\mu_\mathrm{B}$ expected for a free $\mathrm{Dy^{3+}}$ ion. The observed reduction can be attributed to strong crystalline electric field (CEF) effect, directly evidenced for DyNiSb in inelastic neutron scattering experiment \cite{Karla1999}. With increasing temperature, the magnitude of $M(B)$ systematically decreases, and the isotherms tend towards straightly-linear behavior. However, even far above $T_\mathrm{N1}$ they remain slightly curvilinear, likely because of short-range magnetic, CEF and/or Zeeman interactions. 

As displayed in the inset to Fig. \ref{mag}(c), the magnetization measured at 2~K shows a small remanence effect that is compatible with the ZFC/FC bifurcation. Both findings hint at the presence of tiny ferromagnetic component in the AFM magnetic structure of DyNiSb. According to the result of neutron diffraction study of polycrystalline samples, the dysprosium magnetic moments are aligned along the [100] direction with a propagation vector of (1/2, 1/2, 1/2)~\cite{Karla}. Thus, the small hysteresis in $M(B)$ can be attributed to little canting of the magnetic moments within the AFM domains or the contribution due to the AFM domain walls, both being much too small to be seen in the neutron experiment performed.

\begin{figure} [b]
\includegraphics[width= 0.8\linewidth]{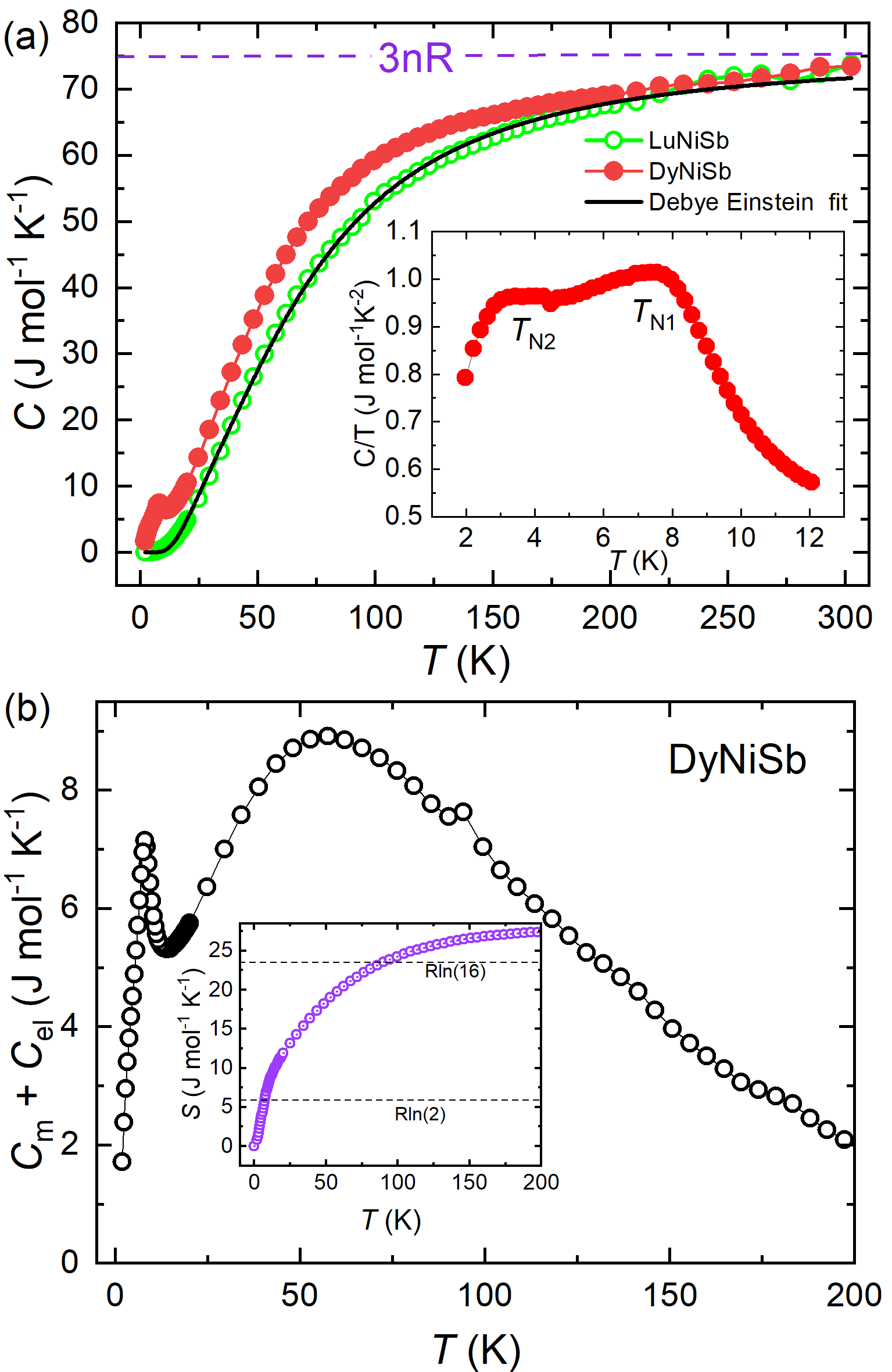}
\caption{(a) Temperature variations of the specific heat of DyNiSb and LuNiSb. The black solid line represents the Debye-Einstein fit described in the text. The inset shows the specific heat over temperature data of DyNiSb in the region of AFM transitions. (b) Temperature dependence of the sum of magnetic and electronic contributions to the specific heat of DyNiSb estimated as described in the text. The inset presents the temperature variation of the magnetic entropy.}
\label{HC}
\end{figure}

Fig.~\ref{HC}(a) presents the temperature dependence of the specific heat, $C(T)$, of DyNiSb, compared with that of its nonmagnetic isostructural counterpart LuNiSb.
Near room temperature, $C(T)$ of both compounds is close to the Dulong - Petit limit (3$nR$ = 74.82 $\mathrm{J \,mol^{-1} K^{-1}}$, where $n$ stands for the number of atoms per formula unit and $R$ is the universal gas constant). Below $\approx\,$200~K, the specific heat of DyNiSb is distinctly larger than that of LuNiSb, signaling the contribution due to splitting the $^6H_{15/2}$ ground multiplet of Dy$^{3+}$ ions in the cubic CEF potential. In the low-temperature region, $C(T)$ exhibits two anomalies at $T_\mathrm{N1}$ and $T_\mathrm{N2}$ corresponding to the AFM transitions revealed in the $\chi(T)$ measurements. While the former singularity has a classical form of a lambda-shaped peak, the other manifests itself only as an inflection in the $C(T)$ curve, and it is better visible in the plot of the ratio $C/T$ versus $T$ (see the inset to Fig.~\ref{HC} (a)).

The specific heat of LuNiSb was analysed in terms of a sum of the lattice, $C_{\rm {ph}}$, and electronic, $C_{\rm {el}} = \gamma T$, contributions ($\gamma$ denotes the Sommerfeld coefficient). The former one was assumed to be describable by a combination of Debye and Einstein models, given by the formula: 

\begin{equation}
\begin{split}
C_{\rm ph}(T) = nR \Bigg[ 
    & 9(1-d) \left( \frac{T}{\Theta_{\rm D}} \right)^3 
    \int_{0}^{x_{\rm D}} 
    \frac{x^4 \exp(x)}{(\exp(x) - 1)^2} dx \\
    & + 3d  \frac{x_{\rm E}^2 \exp(x_{\rm E})}
    {(\exp(x_{\rm E}) - 1)^2} 
\Bigg],
\end{split}
\end{equation}

\noindent where $x_{\rm D}~= \Theta_{\rm D}/T$, $x_{\rm E}~= \Theta_{\rm E}/T$, $\Theta_{\rm D}$ and $\Theta_{\rm E}$ stand for the Debye and Einstein temperature, respectively, and $d$ is a weight factor of the Debye and Einstein contributions. Fitting Eq. 4 to the experimental data of LuNiSb resulted in the $C(T)$ curve shown in Fig.~\ref{HC}(a) by the solid line, and yielded the parameters: $\gamma =$ 3\,mJ\,mol$^{-1}$K$^{-2}$, $\Theta_{\rm D}=$ 345~K, $\Theta_{\rm E}=$ 100~K, and $d =$ 0.33. The so-obtained values are in good agreement with those reported in Ref. \cite{Ciesielski}, where a slightly different approach for modeling $C(T)$ of LuNiSb was adopted. 

Fig.~\ref{HC}(b) presents the sum of electronic and magnetic contribution, $C_{\rm el}+C_{\rm m}(T)$, to the specific heat of DyNiSb, derived by subtracting the lattice specific heat of LuNiSb from the experimental data. The sharp peak at low temperatures marks the AFM transition, while a broad maximum centered around 55~K arises due to the CEF interactions. By integration of the $C_{\rm m}/T(T)$ data one can calculate the magnetic entropy, $S(T)$, which is however hampered by the lack of reliable estimate of $C_{\rm el}$ in DyNiSb. For this reason, the magnitude of the entropy shown in the inset to Fig.~\ref{HC}(b) is overestimated. Near $T_\mathrm{N1}$, the entropy reaches a value close to $R$\,ln2, characteristic of a doublet ground state in Kramer's system. At high temperatures, the magnitude of $S$ exceeds the theoretical limit $R$\,ln16, expected for Dy$^{+3}$ ion with the total angular momentum $J$=15/2. The discrepancy can be attributed to the mentioned shortcomings of the performed analysis of the heat capacity data.

\subsection{Electrical resistivity and magnetotransport}

\begin{figure*} [hbt]
\includegraphics[width=0.9\linewidth]{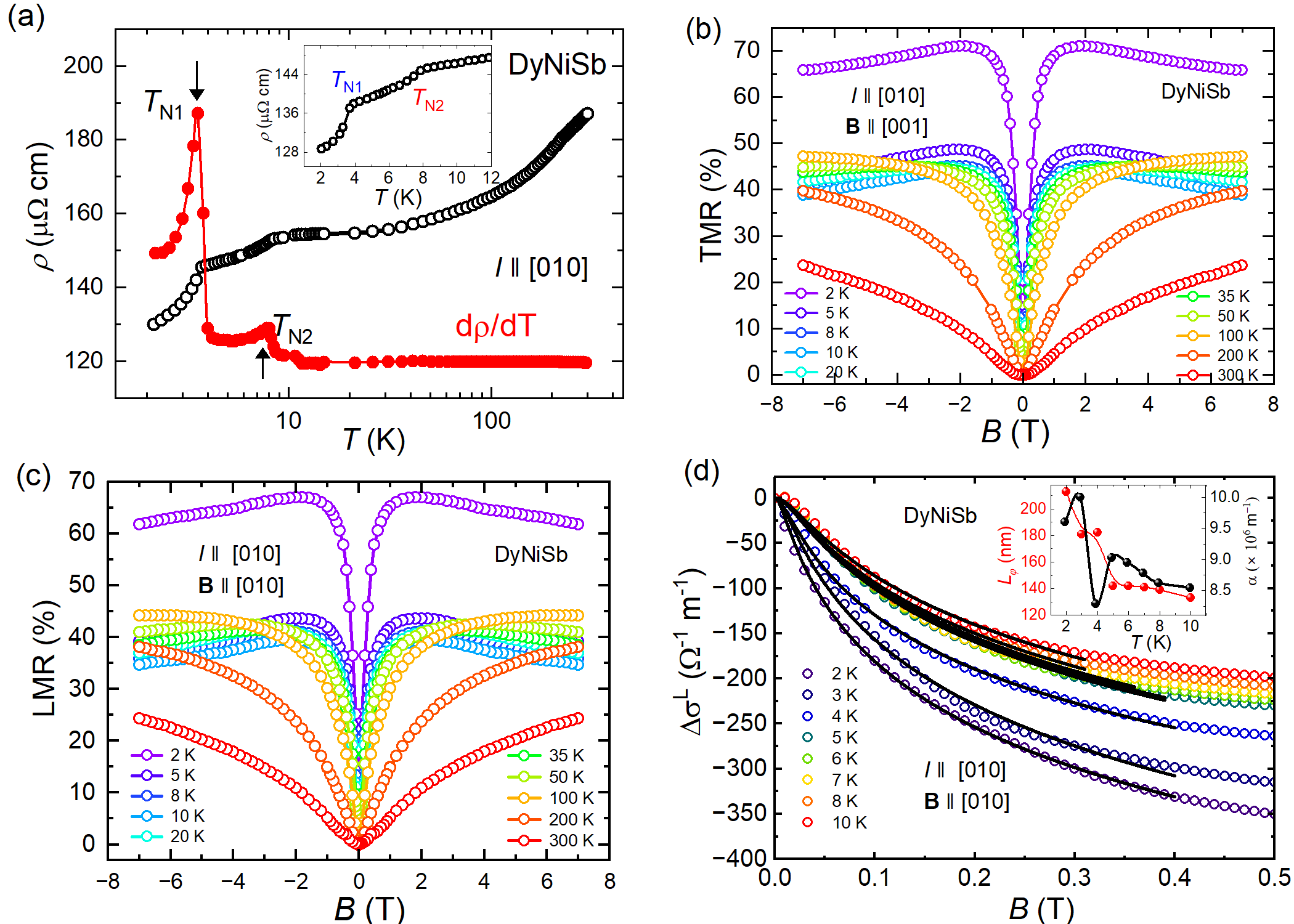}
\caption{(a) Temperature dependence of the electrical resistivity of single-crystalline DyNiSb measured along the crystallographic direction [010] presented on a semi-logarithmic scale. The red symbols represent the temperature derivative of the resistivity (unit-less representation). The inset shows the low-temperature data.  (b) The magnetoresistance isotherms taken for DyNiSb at various temperatures with the current $\textbf{I}$ $\parallel$ [010] and magnetic field aligned along the [001] axis. (c) The magnetoresistance isotherms of DyNiSb measured in parallel configuration of the electric current and magnetic field $\textbf{I}$ $\parallel$ $\textbf{B}$ $\parallel$ [010]. (d) The magnetoconductivity isotherms of DyNiSb calculated from the longitudinal magnetoresistance data collected as in panel (c) at low temperatures and in weak magnetic fields. The solid  lines represent the Hikami-Larkin-Nagaoka fit described in the text. The inset displays the temperature variations of the fitting parameters.}
\label{res}
\end{figure*}

The temperature variation of the electrical resistivity, $\rho(T)$, of single-crystalline DyNiSb, measured with electric current flowing along the crystallographic direction [010], is shown in Fig. \ref {res}(a). At odds with the previous reports \cite{skolozdra1997, karla1998, Pierre2000, Ciesielski2020}, a metallic-like behavior was found, though the observed decrease in the resistivity with decreasing temperature is rather small, hinting at large degree of structural imperfections in the measured specimen. At low temperatures, two distinct anomalies occur in $\rho(T)$, which correspond to the AFM phase transitions seen in the $\chi(T)$ and $C(T)$ data (see the inset to Fig.~\ref{res}(a)).

The results of the resistivity measurements of DyNiSb single crystals carried out in external magnetic fields are shown in Fig. \ref{res}(b) and Fig. \ref{res}(c) in a form of field variations of the magnetoresistance, defined as MR = $[\rho(B)-\rho(0)]/\rho(0)$, for transverse (\textbf{I} $\perp$ \textbf{B}; TMR) and longitudinal (\textbf{I} $\parallel$ \textbf{B}; LMR) configurations, respectively. 
Remarkably, the TMR and LMR isotherms exhibit very similar behavior, being dominated in weak magnetic fields and low temperatures by a strong positive component, while in strong fields by a negative one, which manifests as a negative slope in the MR$(B)$ curves. At 2~K, for both configurations, the positive magnetoresistance attains a value as large as about 70\% near $B =$ 2~T. In the strongest field applied it decreases to 62-65\% depending on the measurement geometry.   

The sharp rise in MR in weak magnetic fields likely originates from weak antilocalization (WAL) effect. The WAL effect is commonly observed for HH compounds \cite{Dan2024, ram2024, Pavlosiuk2019}, and usually interpreted in terms of the Hikami-Larkin-Nagaoka (HLN) approach. Despite the HLN theory was originally developed for thin films, where charge transport is governed mostly by surface states, it is commonly used also for various 3D systems, like HH phases \cite{Pavlosiuk2016}. According to the HLN model, the magnetoconductivity, $\Delta \sigma$, measured in weak magnetic field region is given by the formula

 \begin{equation}
       \Delta\sigma = \alpha\frac{e^2}{2\pi^2\hbar}\left[\Psi\left(\frac{1}{2}+\frac{\hbar}{4eL{_\phi^2}B}\right)-\ln\left(\frac{\hbar}{4eL_{\phi}^2B}\right)\right]
    \label{HLN}
    \end{equation}

\noindent where the prefactor $\alpha$ is related to the strength of spin-orbit coupling, $L_{\phi}$ denotes the phase coherence length, and $\Psi$ stands for the digamma function.

Fig. \ref{res}(d) presents the weak-field magnetoconductivity in DyNiSb, derived as an inverse of the LMR isotherms collected at several low temperatures. As can be inferred from this figure, the experimental data can be well approximated by the HLN function (note the solid lines) with the fitting parameters shown in the inset. At 2~K, $L_{\phi}$ is equal to 213~nm, and decreases with increasing temperature, reflecting the WAL effect weakening. The parameter $\alpha$ is in the order of $\mathrm{10^6 m^{-1}}$, which is typically observed for HH materials \cite{Dan2024, Pavlosiuk2016}. 

\begin{figure} [hbt]
\includegraphics[width=0.9\linewidth]{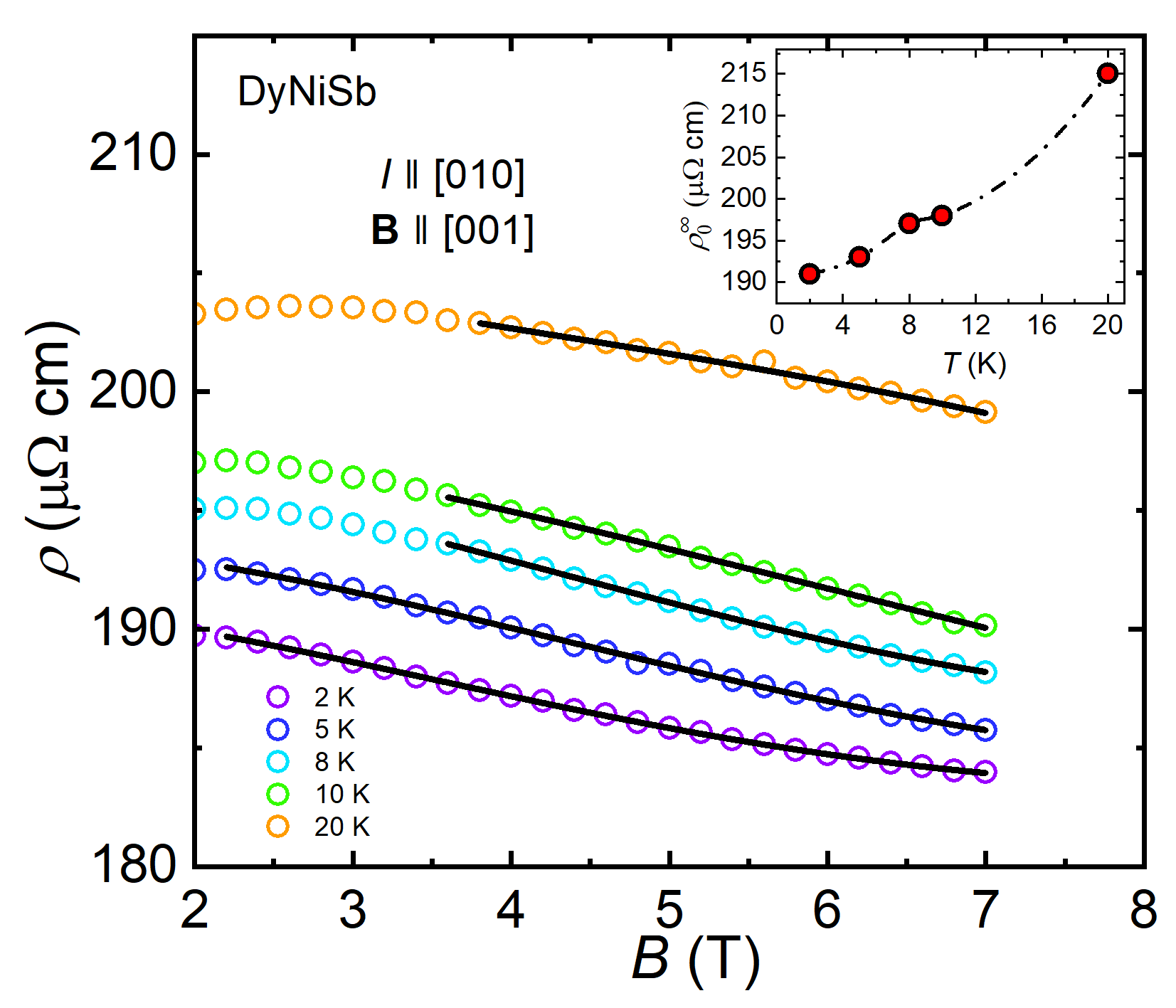}
\caption{ Magnetic field variations of the electrical resistivity of single-crystalline DyNiSb measured at different temperatures with electric current $\textbf{I}$ $\parallel$ [010] and magnetic field aligned along the [001] axis. Solid lines represent the dGF fits discussed in the text. The inset shows the temperature dependence of the parameter $\rho_0^\infty$ derived from the dGF model.}
\label{FIT}
\end{figure}

The negative contribution to MR observed For DyNiSb in strong magnetic fields can be attributed to field-induced polarization of magnetic moments that results in gradual suppression of the spin-disorder scattering. Fig.\ref{FIT} presents the field variations of the electrical resistivity measured along the crystallographic direction [010] in transverse magnetic field applied along the [001] axis. The solid lines represent the least-squares fits of the experimental data in terms of the de Gennes and Friedel (dGF) model:

\begin{equation}\label{dGF}
    \rho_{\rm {dGF}}(B) = \rho_0^\infty  [1-(M(B)/M_s)^2] ,
\end{equation}

\noindent where the parameter $\rho_0^\infty$ denotes the strength of the spin-disorder scattering, and $M_\mathrm{s}$ stands for the saturation magnetization estimated by the approximation of the experimental $M(B)$ data of DyNiSb by the Brillouin function. In the inset to Fig.~\ref{dGF}, there is shown the temperature variation of the parameter $\rho_0^\infty$ derived from the dGF fits. In turn, the values of $M_\mathrm{s}$ were found to be close to 6.5~$\mu_\mathrm{B}$ up to 10 K, and 6.3~$\mu_\mathrm{B}$  at 20 K. These results validate the applicability of the dGF approach, being consistent with the experimentally observed magnitudes of the electrical resistivity and the magnetization of DyNiSb.

With increasing temperature, WAL and spin-polarization effects become less effective, giving rise to a predominance of positive MR, which can be putatively attributed to multiband conduction. It is worth noting that very similar MR response at high temperatures was found before for other HH antimonides, like TmPdSb \cite{Dan2024} or ErPdSb \cite{Agarwal2025}, as well as related hexagonal compounds, e.g., YbAuSb \cite{ram2024}. 

\begin{figure*}[t]
\includegraphics[width= 0.7\linewidth]{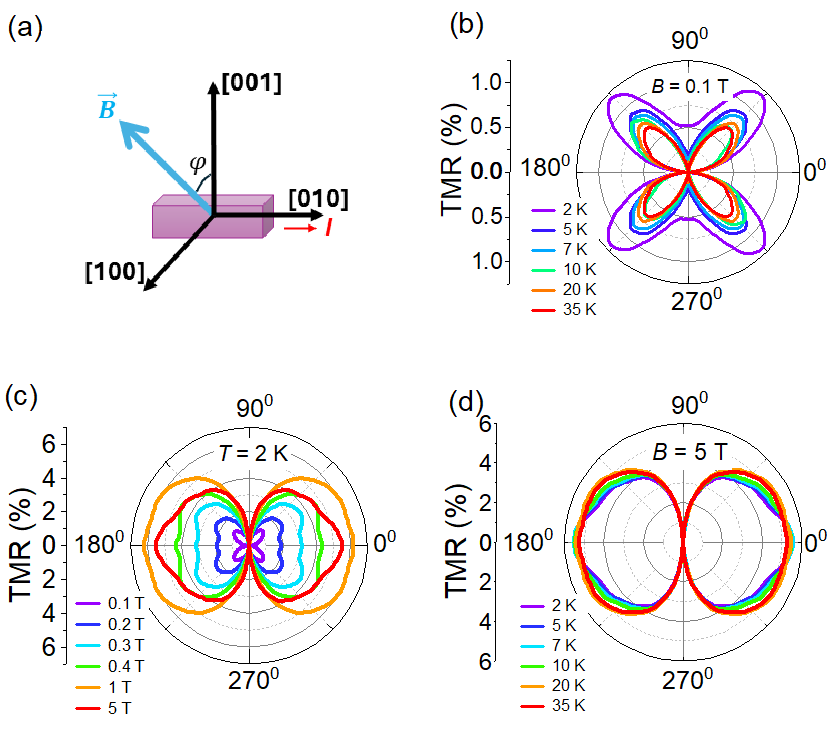}
 \caption{(a) Schematic illustration of the experimental geometry used in the study of angular variation of the transverse magnetoresistance in single-crystalline DyNiSb. (b) Polar plots of the magnetoresistance measured as shown in panel (a) in an external magnetic field of 0.1~T at several different temperatures. (c) Polar plots of the magnetoresistance measured as shown in panel (a) at a temperature of 2~K in several different magnetic fields. (d) Polar plots of the magnetoresistance measured as shown in panel (a) in an external magnetic field of 5~T at several different temperatures.}
\label{amr}
\end{figure*}

 \subsection{Angular dependence of magnetoresistance}

Figure \ref{amr} presents the angular dependencies of the transverse magnetoresistivity of single-crystalline DyNiSb, measured with the current $\textbf{I} \parallel$ [010] and the magnetic field $\textbf{B}$ rotated by angle $\varphi$ in the (010) plane, as depicted in panel (a). The polar plot shown in Fig.\,\ref{amr}(b) displays the data collected in a magnetic field of 0.1~T at several temperatures in the AFM and paramagnetic state. Remarkably, each isotherm bears a four-fold symmetry, and only the magnitude of TMR decreases with increasing temperature. Generally, the character of MR($\varphi$) is governed by the shape of Fermi surfaces, character of magnetic structure, and crystal structure of a given material \cite{pippard1989}. For example, in TbPtBi, the four-fold symmetry of MR($\varphi$) was observed only within the AFM state, attributed to the magnetic unit cell symmetry, whereas in the paramagnetic state, the symmetry was two-fold, reflecting the cubic crystal structure \cite{ chen2023}. In the present case of DyNiSb, the persistence of the four-fold symmetry even at 35~K, well above the transition temperature, suggests that the principal factor determining MR($\varphi$) in weak magnetic fields is the form of Fermi surface. 

 With further increase in magnetic field, MR($\varphi$) measured at $T=2$\,K begins to exhibit both two-fold and four-fold components like that seen already in $B =$ 0.2~T (see Fig.~\ref{amr}(c)). This feature can be considered as a sign of field-induced reconstruction of the Fermi surface, a common phenomenon in HH materials \cite{chen2023, Chen20231}. In a field of 5 T, the two-fold character of MR($\varphi)$) is obvious, and becomes further reinforced with increasing temperature (see Fig.~\ref{amr}(d)).  

\begin{figure}[]
\includegraphics[width= \linewidth]{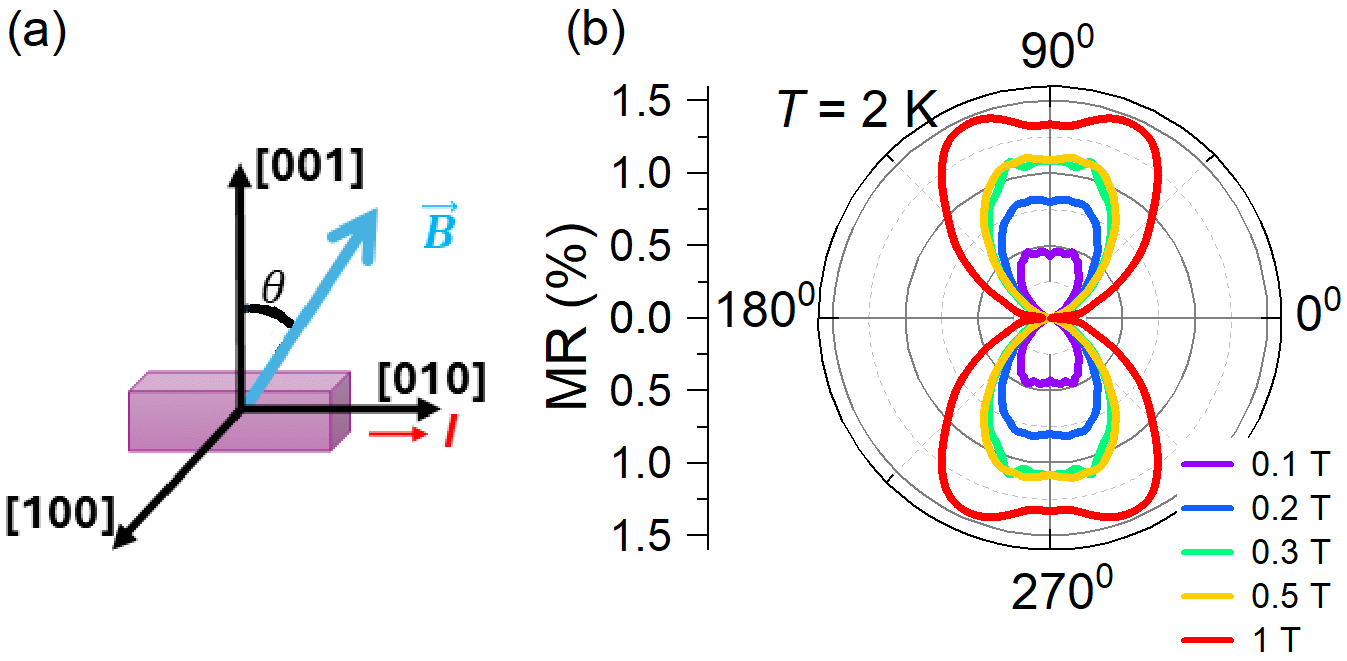}
 \caption{(a) Schematic illustration of the experimental geometry used in the measurements of the angular variation of the magnetoresistance in single-crystalline DyNiSb with field rotation in the (100) plane. (b) Polar plots of the magnetoresistance measured as displayed in panel (a) at a temperature of 2~K in several different magnetic fields.}
\label{pamr}
\end{figure}

In contrast to the hypothetical Fermi surface reconstruction induced be increasing magnetic field applied perpendicular to the electric current direction, no similar effect was found in the MR data of DyNiSb measured in external magnetic field rotated within the (100) plane. As visualized in Fig.~\ref{pamr}, the angular dependence of the magnetoresistance corresponding to the field rotation from [001] to [010] has a clear two-fold symmetry up to at least 1\,T. Similar behavior was reported for other HH compounds \cite{Lu2025}.

\subsection{Defect formation energies and electronic structure}

\begin{figure*}[] 
\includegraphics[width= 0.9\linewidth]{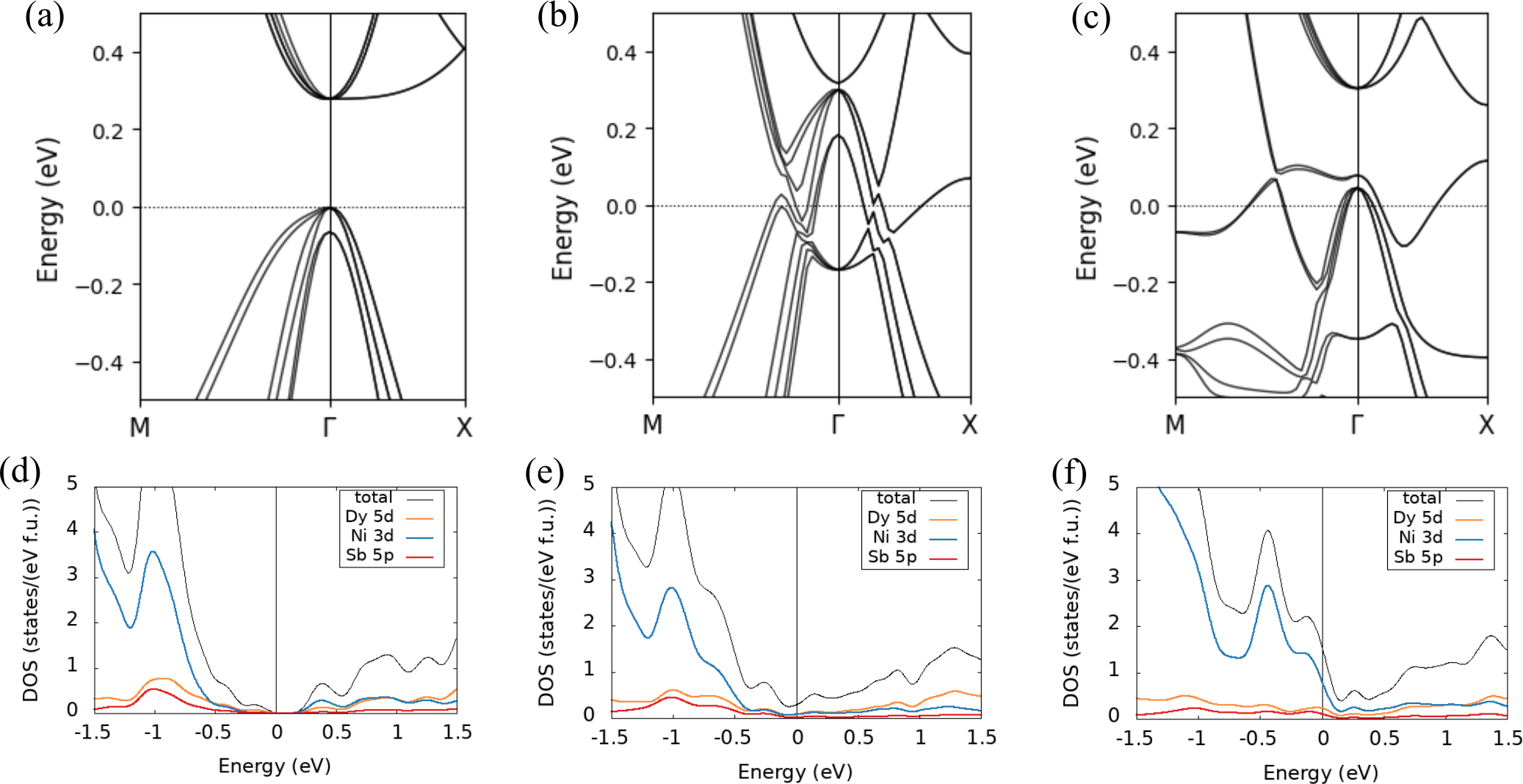}
 \caption{ Band structures calculated for DyNiSb assuming (a) the MgAgAs-type crystal structure with no structural defects, (b) the presence of vacancies at the Ni atom sites, and (c) the presence of extra Ni atoms at intersitial positions. (d-f) Total and partial DOS plots for the structural models considered in panels (a-c)}
\label{DFT_FIG}
\end{figure*}

\begin{figure}
\includegraphics[width= \linewidth]{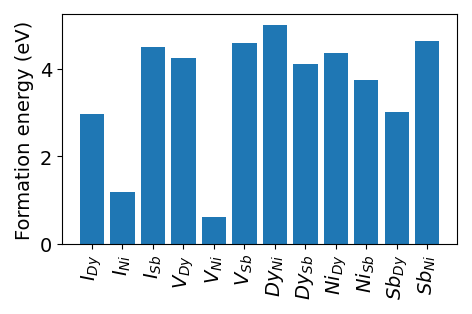}
 \caption{Defect formation energies calculated from first-principles for DyNiSb.}
\label{DFT_FE}
\end{figure}

As presented in Fig.~\ref{DFT_FIG}(a), the ideal DyNiSb system is expected to exhibit a semiconducting character with a narrow band gap of 0.28 eV. The previous electronic band-structure calculations also showed the narrow-gap semiconducting nature of this HH antimonide with similar magnitude of the band gap \cite{Mehta2025, Baidak2023}. However, these results are in disagreement with the experimental data obtained for single-crystalline DyNiSb, which revealed a metallic character of its electronic transport. It is well documented that the HH compounds are notorious for inherent structural defects that may lead to a finite density of states at the Fermi level. To examine this possibility, the values of formation energy for particular possible intrinsic defects in DyNiSb were computed using a DFT-based approach. As can be inferred from Fig. \ref{DFT_FE}, a relatively low value E($V_{Ni}$) = 0.62 eV was obtained for the model assuming Ni vacancies, which is consistent with the experimental findings for a sister compound ScNiSb \cite{Harmening2009}. Another possibility seems the presence of interstitial Ni atoms in the crystallographic unit cell of DyNiSb. Other types of defects are rather unlikely because of the significantly higher formation energies.

Previous theoretical studies revealed that the effect of interstitial defects on the electronic structures of HH semiconductors may be crucial to understanding the strong overestimation of band gaps by the theoretical results obtained for perfect crystal structures with no defects \cite{hH_defect}. The Ni-bearing materials (e.g., ZrNiSn and TiNiSn) are expected to exhibit very low values of E($I_{Ni}$) and narrow band gaps resulting from high defect concentrations \cite{hH_defect}.

As can be inferred from Fig. \ref{DFT_FIG}(b), the presence of vacancies at the Ni atom sites results in smaller Fermi surface sheets, as one may expect considering the effect of interstitial Ni atoms (compare Fig. \ref{DFT_FIG}(c)). According to the density of states (DOS) plots depicted in Fig. \ref{DFT_FIG}(d-e), the total DOS in the valence and conduction band regions in DyNiSb is dominated by the Ni $3d$ states with minor contributions of the Dy $5d$ and Sb $5p$ states. It should also be noted that in the case of $I_{Ni}$, the Fermi level is located on a steep slope of DOS. The key result from the DFT-based modeling is the observation that regardless the type of intrinsic defects their presence in the unit cell of DyNiSb brings about a metallic character of the electrical transport in this compound, in concert with the experimental findings.

\section{CONCLUSIONS}

The antimonide DyNiSb is a representative of the family of cubic half-Heusler phases. The compound was previously investigated on polycrystalline samples only \cite{skolozdra1997, karla1998, Karla1999, Pierre2000, synoradzki2018}. 
The experimental data obtained in the present work on high-quality single crystals of DyNiSb differ significantly from those described in the literature. The thermodynamic and electrical transport measurements revealed two successive antiferromagnetic transitions at $T_\mathrm{{N1}}$ = 7.3~K and $T_\mathrm{{N2}}$ = 3.4~K. The crystals were found to exhibit a metallic behavior, with very similar change of the electrical resistance in external magnetic fields applied along and transverse to the current direction. At low temperatures, the magnetoresistance measured in fields $B <$ 0.5~T is dominated by the weak antilocalizaton effect, while in stronger fields, its magnitude decreases due to increasing spin polarization. At high temperatures ($T\geq200$ K), a positive unsaturated magnetoresistance was observed. Angle-dependent magnetoresistance measurements carried out with $\textbf{I} \parallel$ [010] and $\textbf{B}$ rotating in the (010) plane, showed a four-fold symmetry in weak magnetic fields, which gradually evolved into two-fold symmetry with increasing field strength, suggesting a field-induced reconstruction of the Fermi surface. In another configuration, with $\textbf{I} \parallel$ [010] and $\textbf{B}$ rotating in the (100) plane, the magnetoresistance exhibited two-fold symmetry regardless of the field strength.

The DFT-based calculations revealed for DyNiSb relatively small formation energies for intrinsic Ni vacancies and Ni interstitial point defects in the MgAgAs-type unit cell. The electronic structures derived with considering such defects revealed a semimetallic character in accordance with the experimental results. The band structure calculated assuming Ni vacancies, having small hole FS pockets, seems much more prone to the reconstruction (e.g. disappearance of these pockets) due to small shift of Fermi level caused by magnetic field. Considering the angular dependence of the transverse magnetoresistance in DyNiSb, and the magnetic field controlled change of its symmetry, the structural model including the Ni- vacancies seems more appropriate for real single-crystalline DyNiSb.

\begin{acknowledgments}
This work was supported by the National Science Centre (NCN, Poland) under project no. 2021/40/Q/ST5/00066. The DFT calculations were performed at the Wroclaw Center for Networking and Supercomputing (project no. 158).
\end{acknowledgments}

\section*{DATA AVAILABILITY}
The data are not publicly available. The data are available from the authors upon reasonable request.

\bibliography{mybib}

\end{document}